# X-ray Scintillation in Lead Halide Perovskite Crystals

M. D. Birowosuto,[1,2,*] D. Cortecchia,[3,4] W. Drozdowski,[5] K. Brylew,[5] W. Lachmanski,[5] A. Bruno,[4] C. Soci[2,4,6,*]

[1]CINTRA UMI CNRS/NTU/THALES 3288, Research Techno Plaza, 50 Nanyang Drive, Border X Block, Level 6, Singapore 637553

[2]Center of Disruptive Photonic Technologies, TPI, SPMS, Nanyang Technological University, 21 Nanyang Link, Singapore 637371

[3]Interdisciplinary Graduate School, Nanyang Technological University, Singapore 639798

[4] Energy Research Institute @ NTU (ERI@N), Research Techno Plaza, Nanyang Technological University, 50 Nanyang Drive, Singapore 637553

[5]Institute of Physics, Faculty of Physics, Astronomy, and Informatics, Nicolaus Copernicus University, Grudziadzka 5, 87-100 Torun, Poland

[6] School of Physical and Mathematical Sciences, Division of Physics and Applied Physics, Nanyang Technological University, 21 Nanyang Link, Singapore 637371

*Corresponding authors: mbirowosuto@ntu.edu.sg; csoci@ntu.edu.sg

**Abstract**

Current technologies for X-ray detection rely on scintillation from expensive inorganic crystals grown at high-temperature, which so far has hindered the development of large-area scintillator arrays. Thanks to the presence of heavy atoms, solution-grown hybrid lead halide perovskite single crystals exhibit short X-ray absorption length and excellent detection efficiency. Here we compare X-ray scintillator characteristics of three-dimensional (3D) $MAPbI_3$ and $MAPbBr_3$ and two-dimensional (2D) $(EDBE)PbCl_4$ hybrid perovskite crystals. X-ray excited thermoluminescence measurements indicate the absence of deep traps and a very small density of shallow trap states, which lessens after-glow effects. All perovskite single crystals exhibit high X-ray excited luminescence yields of >120,000 photons/MeV at low temperature. Although thermal quenching is significant at room temperature, the large exciton binding energy of 2D $(EDBE)PbCl_4$ significantly reduces thermal effects compared to 3D perovskites, and moderate light yield of 9,000 photons/MeV can be achieved even at room temperature. This highlights the potential of 2D metal halide perovskites for large-area and low-cost scintillator devices for medical, security and scientific applications.

**Introduction**

The investigation of X-ray detectors started with the discovery of X-rays by Wilhelm Röntgen, who noticed the glow from a barium platino-cyanide screen placed besides a vacuum tube [1,2]. Since this discovery, more than one hundred years ago, the development of efficient [3-5] and large-area [5-7] X-ray detectors has been a topic of continuous interest, targeting a wide range of applications, from crystallography [8] to space exploration [9].

Modern X-ray detectors rely on two main mechanisms of energy conversion. The first is photon-to-current conversion, in which a semiconducting material directly converts the incoming radiation into electrical current [4-6]; the second is X-ray to UV-visible photon down-conversion, in which a scintillator material is coupled to a sensitive photodetector operating at lower photon energies [2]. Both methods are equally compelling for practical implementations, although their viability will ultimately depend on the development of new materials to overcome some of the current limitations, such as high cost, small area, and low conversion efficiency of the X-ray absorbers. Recent demonstrations of the use of hybrid metal-halide perovskites for X- and γ-ray detection has spurred great interest in this class of materials [7,10,11,12]. Besides their good detection efficiency, solution processing holds great promise for facile integration and development of industrial and biomedical applications.

Methylammonium lead trihalide perovskites ($MAPbX_3$ where $MA=CH_3NH_3$ and $X=$ I, Br, or Cl) have demonstrated excellent performance in optoelectronic devices like field effect transistors [13], highly sensitive photodetectors for visible region [14], and light emitting devices [15,16]. Moreover, compositional tuning was used to realize tunable-wavelength lasers [17]. As X-ray detectors, $MAPbX_3$ yield notably large X-ray absorption cross section due to large atomic numbers of the heavy Pb and I, Br, Cl atoms [10,11]. Thin-film $MAPbX_3$ p-i-n photodiode and lateral photoconductor devices have shown good efficiency for X-ray photon-to-current conversion [10,11]. However, thin-film X-ray detectors have typically low responsivity at high (keV) photon energies, where the absorption length (~mm) is much larger than the film thickness (~$\mu$m); even if thickness is increased to improve detection probability, direct photon-to-current conversion is ultimately hampered by the limited carrier-diffusion length (~1 $\mu$m in perovskites) [10]. Efficient X-ray photon-to-current conversion has been shown recently in single-crystal (thick) perovskite MAPbBr$_3$, but sensitivity is still limited to energies up to 50 keV [11]. Also, standard $\gamma$-photon counting for energies up to 662 keV has been demonstrated in MAPbI$_3$ [12].

As opposed to direct photon-to-current conversion detectors, X-ray scintillators do not suffer from limited carrier diffusion length of the absorbing material [18,19]. Thin films of phenethylammonium lead bromide, PhE-PbBr$_4$, with sub-nanosecond scintillation decay time have been previously tested in X-ray [20] and proton [21] scintillators, but yielded only 5-6% detection efficiency of 60 keV X-rays, limited by the film thickness (200 $\mu$m) [21]. By combining the good high-energy response with large absorption cross section deriving from large thickness and high mass-density, single crystal perovskite scintillators are therefore expected to improve detection efficiency of keV X- or $\gamma$-rays.

In this paper, we present a thorough comparative study of the scintillation properties of three-dimensional (3D) and two-dimensional (2D) low-bandgap perovskite single crystals. We have synthesized mm-scale 3D perovskite crystals MAPbI$_3$ and MAPbBr$_3$, and 2D perovskite crystal (EDBE)PbCl$_4$ (EDBE=2,2'-(ethylenedioxy)bis(ethylammonium)), comprising of alternating organic and inorganic layers which form a multi-quantum-well-like structure. The excellent quality of these crystals is indicated by structural analysis and by the very small density of shallow traps ($n_0$~$10^5$-$10^7$ cm$^{-3}$, $E$~10-90 meV) determined by X-ray excited thermoluminescence, which reduces after-glow effects. Thanks to their lower bandgap compared to traditional scintillator crystals [6], perovskite crystals produce extremely high light yields of >120,000 photons/MeV (as estimated from X-ray-excited luminescence) at low temperature. In 3D perovskites, the light yield is greatly reduced at room temperature (<1,000 photons/MeV) due to strong thermal quenching effects. Conversely, the 2D perovskite crystal is far more robust against thermal quenching thanks to its large exciton binding energy (~360 meV) induced by charge confinement within the inorganic layers. These results confirm the excellent properties of metal-halide perovskites for X-ray detection, and highlight the potential of 2D perovskite crystals with large exciton binding energy for high-light yield X-ray scintillators.

**Results and discussion**

To study scintillation performance, we have synthesized the high-quality, large-size (~30 to 100 mm$^3$) perovskite single crystals shown in Fig. 1 (see *Materials and methods* section for details on crystal growth). MAPbX$_3$ (X=I, Br) crystals have the conventional three dimensional ABX$_3$ perovskite structure, consisting of a continuous network of corner sharing PbX$_6^{4-}$ octahedra with methyl-ammonium cations occupying the interstitial sites [22,23]. XRD patterns of the ground crystals confirm the formation of the desired perovskites MAPbBr$_3$ and MAPbI$_3$, having cubic and tetragonal crystal structure, respectively (see Supplementary Figs. S1 and S2). Conversely, (EDBE)PbCl$_4$ belongs to the general class of APbX$_4$ (X = I, Br, Cl and A=bidentate organic cation) "two-dimensional" perovskite crystals [24]; it consists of the stack of <100>-oriented perovskite inorganic layers forming a 2D Pb-X network in alternation with organic sheets of di-ammonium cations EDBE$^{2+}$ (Fig. 1). The presence of pronounced 001 and higher order 00l reflections in the XRD pattern indicates unequivocally the formation of the layered perovskite with monoclinic crystal structure (see Supplementary Fig. S3). To the naked eye MAPbI$_3$,

MAPbBr$_3$, and (EDBE)PbCl$_4$ crystals appear lustrous black, orange, and white, respectively. The corresponding glows under ultraviolet lamp excitation are green and white for MAPbBr$_3$ and (EDBE)PbCl$_4$ crystals, while the glow of MAPbI$_3$ could not be observed since its emission lies in the near infrared. Crystal colors and glows agree well with the absorption and photoluminescence properties of the corresponding thin films, which show optical energy gaps of $E_g$=1.51, 2.18, and 3.45 eV for MAPbI$_3$ [22], MAPbBr$_3$ and (EDBE)PbCl$_4$ [24], respectively (Supplementary Fig. S4).

Perovskite crystals offer multiple advantages for X-ray scintillation, specifically: i. Since the light yield of X-ray scintillation is inversely proportional to the optical bandgap $E_g$ [2,18], low-bandgap perovskites of MAPbI$_3$, MAPbBr$_3$ and (EDBE)PbCl$_4$ are expected to yield up to about 270,000, 190,000, and 120,000 photons/MeV, respectively. Those light yields are much higher than state-of-art cerium (Ce$^{3+}$) doped lanthanum tribromides LaBr$_3$ ($E_g$=5.90 eV) [25,26] and Ce$^{3+}$ doped lutetium iodides LuI$_3$ ($E_g$=4.15 eV) [27,28] scintillators, with light yields of 68,000 and 100,000 photons/MeV, respectively. ii. Since X-ray absorption length scales with the effective atomic number $Z_{eff}$ and mass density $\rho$ [2], MAPbI$_3$, MAPbBr$_3$, and (EDBE)PbCl$_4$ ($Z_{eff}$=66.83, 67.13, and 67.52, $\rho$=3.947, 3.582, and 2.191 gr/cm$^3$, respectively) should reach X-ray absorption lengths up to 1 cm at 1 MeV, similar to Ce$^{3+}$-doped LaBr$_3$ and LuI$_3$ scintillators (see Supplementary Fig. S5). iii. The unusually large Stokes shift of two-dimensional (EDBE)PbCl$_4$ could be particularly beneficial to the scintillation yield [29], which is substantially reduced by self-absorption of the luminescence [30]. iv. The extremely fast photoluminescence decay of MAPbI$_3$, MAPbBr$_3$, and (EDBE)PbCl$_4$ (fast components of 4.3, 0.8-5.2, and 7.9 ns, respectively) may provide faster scintillation than 15 ns of Ce$^{3+}$-doped LaBr$_3$ [25,30] and 33 ns of Ce$^{3+}$-doped LuI$_3$ [28] (Supplementary Fig. S6). Nanosecond scintillation decay times were indeed demonstrated in PhE-PbBr$_4$ using X-ray and γ-ray pulses, consistent with time-resolved photoluminescence [20,31]. v. Finally, long emission wavelengths in the range of 400 to 700 nm allow optimal detection of scintillation using highly sensitive avalanche photodiodes (APDs), which can reach quantum efficiencies up to 90-100% in comparison with photomultipliers (PMTs) with only 40-50% efficiency [28].

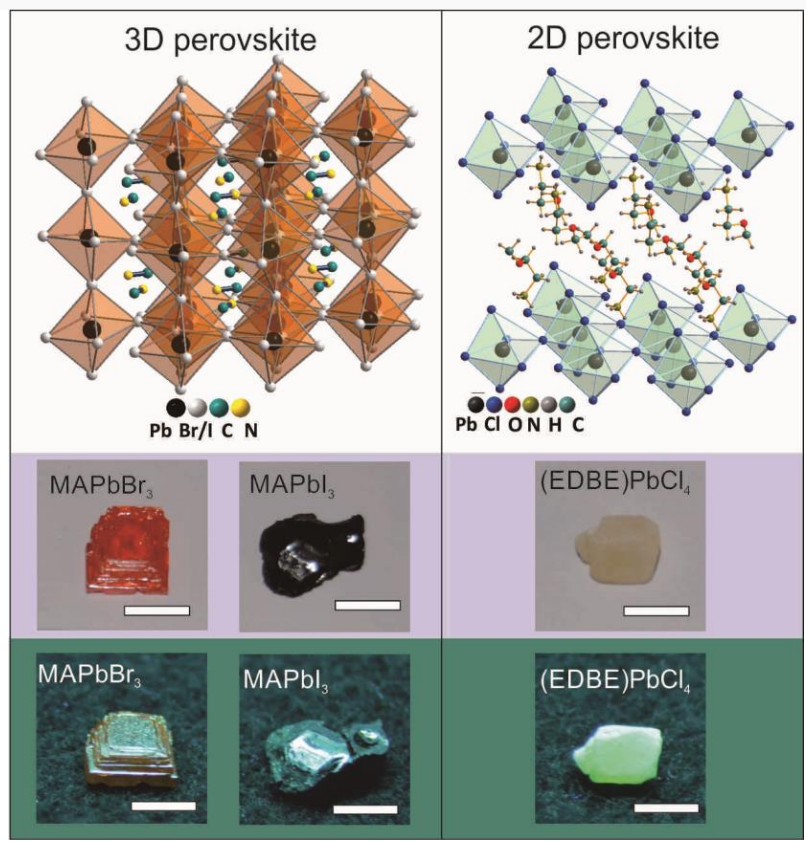

**Figure 1. Crystal structure and appearance.** Top row: crystal structure representation of MAPbX$_3$ (X=I, Br) three-dimensional perovskites (left), and (EDBE)PbCl$_4$ two-dimensional perovskite (right); Middle row: photographs of the large single crystals of hybrid lead halide perovskites; Bottom row: glow of the crystals under ultraviolet lamp excitation. Scale bars: 5 mm.

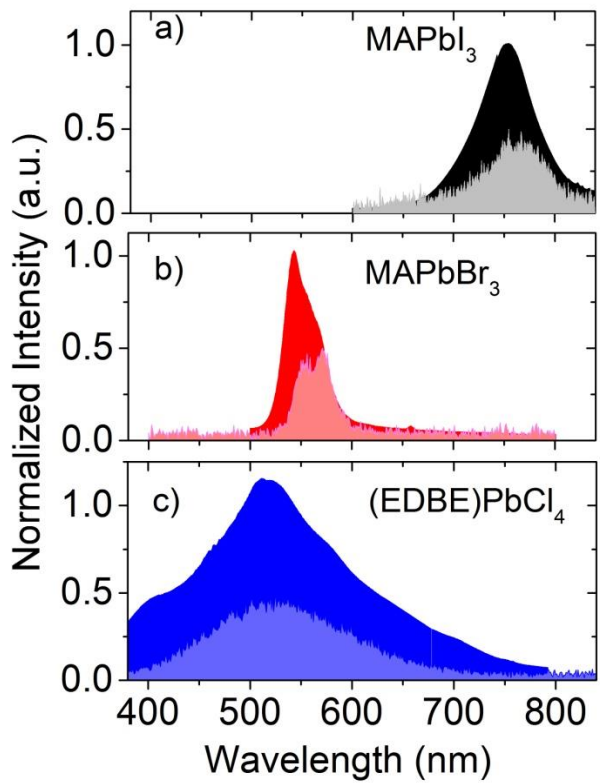

**Figure 2. Emission spectra under X-ray and optical excitation.** X-ray excited luminescence (light color area) and photoluminescence (dark color area) spectra of a) MAPbI$_3$, b) MAPbBr$_3$, and c) (EDBE)PbCl$_4$ recorded at room temperature with excitation wavelengths for photoluminescence of 425, 500, and 330 nm, respectively. Photoluminescence and X-ray excited luminescence spectra were normalized to their maxima, and normalized X-ray excited luminescence spectra were divided by a factor of two for clarity.

The X-ray excited luminescence and photoluminescence spectra of MAPbI$_3$, MAPbBr$_3$, and (EDBE)PbCl$_4$ crystals recorded at room temperature are shown in Fig. 2 (see experimental details in *Materials and methods* section). Both X-ray excited luminescence and photoluminescence spectra of MAPbI$_3$ have a single broadband peak centered at 750 nm with FWHM of ~80 nm (Fig. 2a). For MAPbBr$_3$, both X-ray excited luminescence and photoluminescence spectra exhibit double peaks centered around 560 and 550 nm, respectively. MAPbBr$_3$ has the narrowest emission band with full width of half-maximum (FWHM) of ~40 nm (Fig. 2b). On the other hand, (EDBE)PbCl$_4$ has the broadest emission band centered at 520 nm, with FWHM of ~160 nm (Fig. 2c). Based on emission wavelength, MAPbBr$_3$ and (EDBE)PbCl$_4$ appear to be the most promising candidates for the scintillators coupled to APD [28].

In all perovskite crystals, X-ray excited luminescence and photoluminescence spectra are very similar, indicating that the dominant scintillation mechanism is straightforward: upon X-ray absorption, high-energy excitations thermalize through ionizations and excitations of atoms, until excitons are generated at energies near the bandgap. X-ray excited luminescence stems solely from the intrinsic excitonic emission of the perovskites, and no other defect states seem to be involved in the scintillation process.

The dynamics of radiative processes in materials under high-energy excitation is often complicated by slower non-exponential components due to charge carrier trapping and re-trapping, which manifest themselves as delayed luminescence, or afterglow. Upon termination of the X-ray excitation, afterglow effects would typically contribute a residual luminescence background with characteristic lifetime of few ms, thus lowering the effective light yield and worsening the signal-to-noise ratio. Afterglow effects are particularly detrimental for applications like computed tomography, in which temporal crosstalk considerably reduces the image quality [2]. Charge carrier trapping and re-trapping processes can be monitored by thermoluminescence measurements. In our specific mode of operation for thermoluminescence intensity measurements, we were able to record steady-state X-ray excited luminescence intensity during irradiation, immediately prior to the thermoluminescence scan (see details in *Materials and methods*). In this way, two distinct integrated intensities can be evaluated: the first one, which we denote as $I_{TL}$, comprising the range from the end of X-ray irradiation till the end of the entire run, while the second one, denoted as $I_{TL}+I_{ssXL}$, comprising the range from the start of the X-ray irradiation until the end of the run. This allows calculating, for each sample, the $I_{TL}/(I_{TL}+I_{ssXL})$ ratio, which can be interpreted as the fraction of the total excitation energy accumulated into traps [19,32]. The value of this ratio, therefore, provides a qualitative estimate of the influence of traps on the scintillation yield.

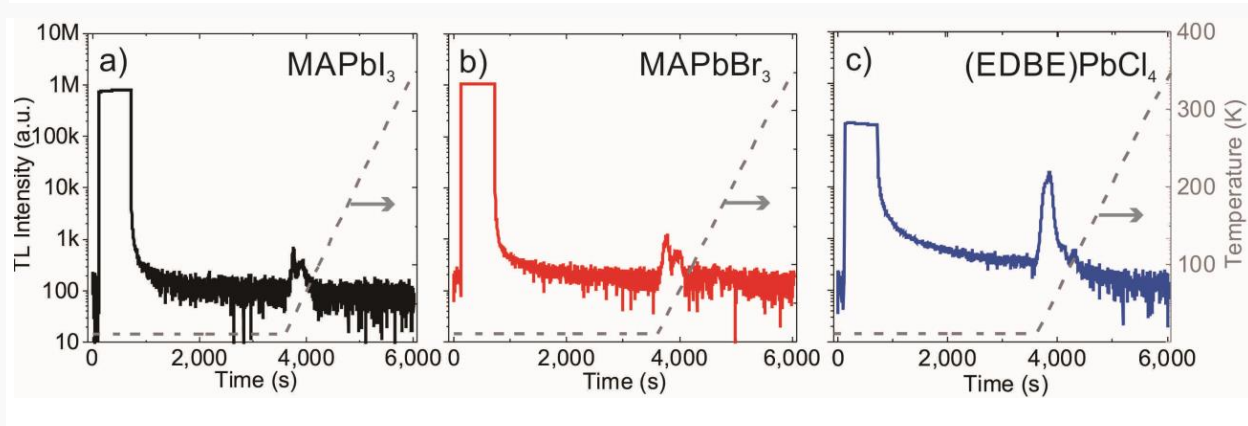

**Figure 3. Residual luminescence background after X-ray excitation.** Low temperature thermoluminescence curves of a) MAPbI$_3$, b) MAPbBr$_3$, and c) (EDBE)PbCl$_4$. The data are presented on a time scale starting at temperature of 10 K and increasing to 350 K after 3600 s, as indicated by the dashed line in the right panel (temperature scale on the right axes).

Typical thermoluminescence curves of the metal halide perovskite crystals are shown as solid curves in Fig. 3. After termination of the X-ray excitation at 10 K, long tails extending to thousands of seconds were observed in all crystals. Although the long-lived component of this afterglow effect is much slower than the photoluminescence decay (see Supplementary Fig. S6), it only occurs at low temperatures (~10 K) and is negligible at room temperature. In the case of MAPbI$_3$ and MAPbBr$_3$, low temperature thermoluminescence curves are dominated by a double-structured peak, with two smaller satellite peaks appearing at longer times (Figs. 3a and 3b). In (EDBE)PbCl$_4$, the low-temperature thermoluminescence curve shows that one peak strongly dominates the other peak while the total intensity of the peaks is much higher than those in MAPbI$_3$ and MAPbBr$_3$ (Fig. 3c). The ratio of $I_{TL}/(I_{TL}+I_{ssXL})$ ~ 0.002 is very similar in both MAPbI$_3$ and MAPbBr$_3$, which is extremely low in comparison with other oxide materials used for scintillators, such as lanthanide aluminium perovskite or garnets [19,32,33,35]. Moreover, MAPbI$_3$ and MAPbBr$_3$ crystals show nearly trap-free behavior from T=75 K up to the highest temperature investigated of T=350 K, a very desirable characteristic from the point of view of scintillation speed and efficiency. In (EDBE)PbCl$_4$, $I_{TL}/(I_{TL}+I_{ssXL})$ ~ 0.058, a slightly higher value than in the three-dimensional perovskite crystals, but still relatively low.

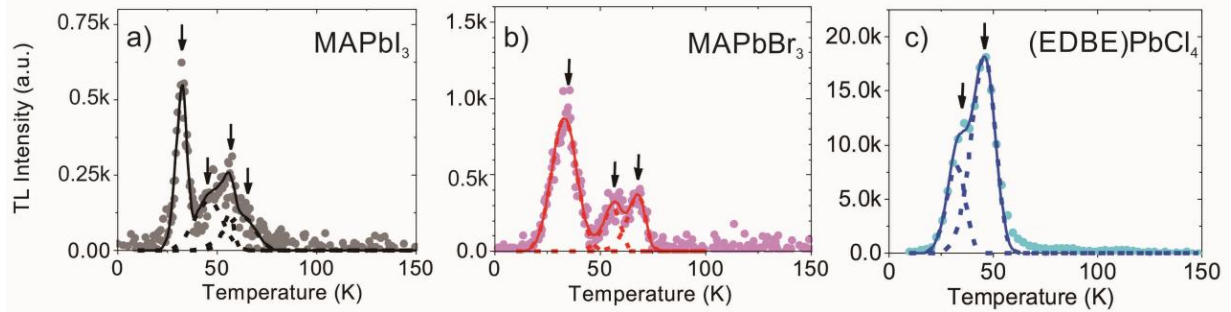

**Figure 4. Determination of low-energy trap states.** Glow curves of a) MAPbI$_3$, b) MAPbBr$_3$, and c) (EDBE)PbCl$_4$ recorded after 10 min X-ray irradiation at 10 K, at a heating rate of 0.14 K/s. The solid lines are the best fits to the experimental data points by multiple Randall-Wilkins equations (Eq. 1): single components and peak temperatures ($T_{max}$) are indicated by dashed lines and arrows, respectively (see Table 1 for fitting parameters).

The zero-order glow curves of the three crystals are presented in Fig. 4. Appearance of thermoluminescence signal at temperatures below 150 K reveals the existence of low-energy trap states. Since for such states it is difficult to determine the exact number of traps, their depth and frequency factors [33], we restrict our analysis to thermoluminescence peaks with intensity larger than 10 % of the maximum. Thermoluminescence curves have been deconvoluted into $k$ glow peaks, based on the classic Randall-Wilkins equation [34]:

$$I_{TL} = \sum_{i=1}^{k} n_{0i} V \sigma_i \exp\left(-\frac{E_i}{k_B T}\right) \exp\left(-\frac{\sigma_i}{\beta} \int_{T_0}^{T} \exp\left(-\frac{E_i}{k_B T'}\right) dT'\right) \qquad (1)$$

where $T$ is the temperature, $\beta$ the heating rate, and $k_B$ the Boltzmann constant; $n_{0i}$ is the initial trap concentration, $V$ is the crystal volume, $E_i$ the trap depth, $\sigma_i$ the frequency factor of each component. Note that the unitless number of traps $n_{0i}V$ is often used to compare the afterglow of different crystals [19,32,33,34,35].

This analysis provides a good indication of the characteristics of prevailing trap states, however it cannot resolve the existence of traps that fall at times much longer than seconds, or with mixed order kinetics [32]. The room temperature lifetime of trapped carriers, such as electron or hole centers and excitons, $\tau_i$, can also be estimated from the energy and frequency factor of the trap, using the well-known Arrhenius formula:

$$\frac{1}{\tau_i} = \sigma_i \exp\left(-\frac{E_i}{k_B T}\right) \qquad (2)$$

While the glow curves of MAPbI$_3$ and MAPbBr$_3$ in Fig. 4a and b have been fitted using four and three components, respectively, the glow curve of (EDBE)PbCL$_4$ in Fig. 4c could be fitted with only two components. The corresponding fitting parameters are shown in Table 1. All crystals have relatively low trap densities, with depth energy ($E$) varying from ~10 to 90 meV. The initial trap concentrations $n_0$ in MAPbI$_3$ and MAPbBr$_3$ can be calculated from the total number of traps ($n_0V$~10$^3$-10$^4$) and the volume of the crystal ($V$~30-100 mm$^3$). The resulting trap concentrations ($n_0$~10$^5$-10$^7$ cm$^{-3}$) are comparable to those of shallow traps previously observed in photoconductivity measurements (~10$^5$-10$^7$ cm$^{-3}$) [11] and space-charge-limited-current (~10$^9$-10$^{10}$ cm$^{-3}$) [23], also considering the uncertainty in the estimate of the active crystal volume. The fastest room temperature lifetimes ($\tau$) of MAPbI$_3$ and MAPbBr$_3$ are of the order of milliseconds, long enough to contribute to the light yield components without residual luminescence background. Correspondingly, logarithmic frequency factors (ln $\sigma$) are all below 16, which is much smaller than ln $\sigma$~30 typically found in pristine or activated oxide materials [19,32,33,35], also reported in Table 1 for comparison. (EDBE)PbCl$_4$ has the largest trap density among the perovskites we investigated, $n_0$~10$^7$ cm$^{-3}$. Large concentration of shallow traps may be beneficial for X-ray scintillation at low-temperature, as in the case of Ce$^{3+}$-doped YAlO$_3$ and LuAlO$_3$ [35], or pristine Li$_2$B$_4$O$_7$ [36]. This is indeed seen in temperature dependent X-ray excited luminescence spectral maps shown in Fig. 5.

| Compound | $T_{max}$ (K) | $E$ (eV) | $\ln \sigma$ (s$^{-1}$) | $\tau$ (s) | $n_0 V$ | Reference |
|---|---|---|---|---|---|---|
| MAPbI$_3$ | 32 | 0.0309 | 8.09 | 1.04·10$^{-3}$ | 2.45·10$^4$ | This work |
|  | 46 | 0.0226 | 1.78 | 0.41 | 1.85·10$^4$ |  |
|  | 56 | 0.0901 | 15.60 | 5.95·10$^{-4}$ | 6.12·10$^3$ |  |
|  | 62 | 0.0389 | 3.25 | 0.18 | 1.45·10$^4$ |  |
| MAPbBr$_3$ | 33 | 0.0139 | 1.16 | 0.54 | 7.61·10$^4$ | This work |
|  | 56 | 0.0602 | 9.02 | 1.31·10$^{-3}$ | 2.10·10$^4$ |  |
|  | 68 | 0.0909 | 12.1 | 2.04·10$^{-4}$ | 2.73·10$^4$ |  |
| EDBEPbCl$_4$ | 32 | 0.0177 | 2.83 | 0.12 | 5.95·10$^5$ | This work |
|  | 45 | 0.0281 | 3.40 | 0.10 | 1.71·10$^6$ |  |
| LuAlO$_3$: Ce$^{3+}$ | 36 | 0.0148 | 0.9346 | 2.16·10$^{-2}$ | 2.84·10$^4$ | [19,32,35] |
|  | 88 | 0.105 | 10.07 | 2.29·10$^{-2}$ | 1.53·10$^4$ |  |
|  | 187 | 0.498 | 27.22 | 2.51·10$^{-2}$ | 2.10·10$^6$ |  |
|  | 206 | 0.385 | 17.56 | 1.61·10$^{-2}$ | 4.64·10$^4$ |  |
|  | 223 | 0.669 | 30.99 | 2.18·10$^{-2}$ | 1.38·10$^4$ |  |
|  | 253 | 0.75 | 30.53 | 2.08·10$^{-2}$ | 4.84·10$^4$ |  |
|  | 273 | 0.799 | 30.08 | 2.05·10$^{-2}$ | 1.52·10$^5$ |  |
| YAlO$_3$: Ce$^{3+}$ | 108 | 0.30 | 29.24 | 4.99·10$^{-2}$ | ~10$^5$ | [19,32,35] |
|  | 154 | 0.50 | 34.18 | 3.02·10$^{-2}$ | ~10$^5$ |  |
|  | 281 | 0.421 | 18.05 | 1.95 | ~10$^4$ |  |
| Gd$_3$Al$_2$Ga$_3$O$_{12}$: Ce$^{3+}$ | 36 | 0.0576 | 15.9 | 1.2·10$^{-6}$ | 1.6·10$^5$ | [33] |
|  | 45 | 0.0446 | 8.32 | 1.4·10$^{-3}$ | 4.7·10$^5$ |  |
|  | 73 | 0.116 | 15.1 | 2.7·10$^{-5}$ | 3.4·10$^5$ |  |
|  | 181 | 0.211 | 9.01 | 0.52 | 2.9·10$^5$ |  |
|  | 240 | 0.527 | 21.31 | 0.65 | 2·10$^5$ |  |
|  | 255 | 0.321 | 9.76 | 19 | 7.5·10$^5$ |  |

**Table 1. Trap state parameters.** The parameters were derived from the fitting of first-order glow curves in Figure 4: $T_{max}$ is the temperature at which the glow curve peaks, $E$ the trap depth, $\ln \sigma$ the logarithmic frequency factor in $s^{-1}$, $\tau$ the room temperature lifetime, and $n_0V$ the total, initial number of traps. Comparative parameters of known scintillator materials from the literature are also reported in the last three lines.

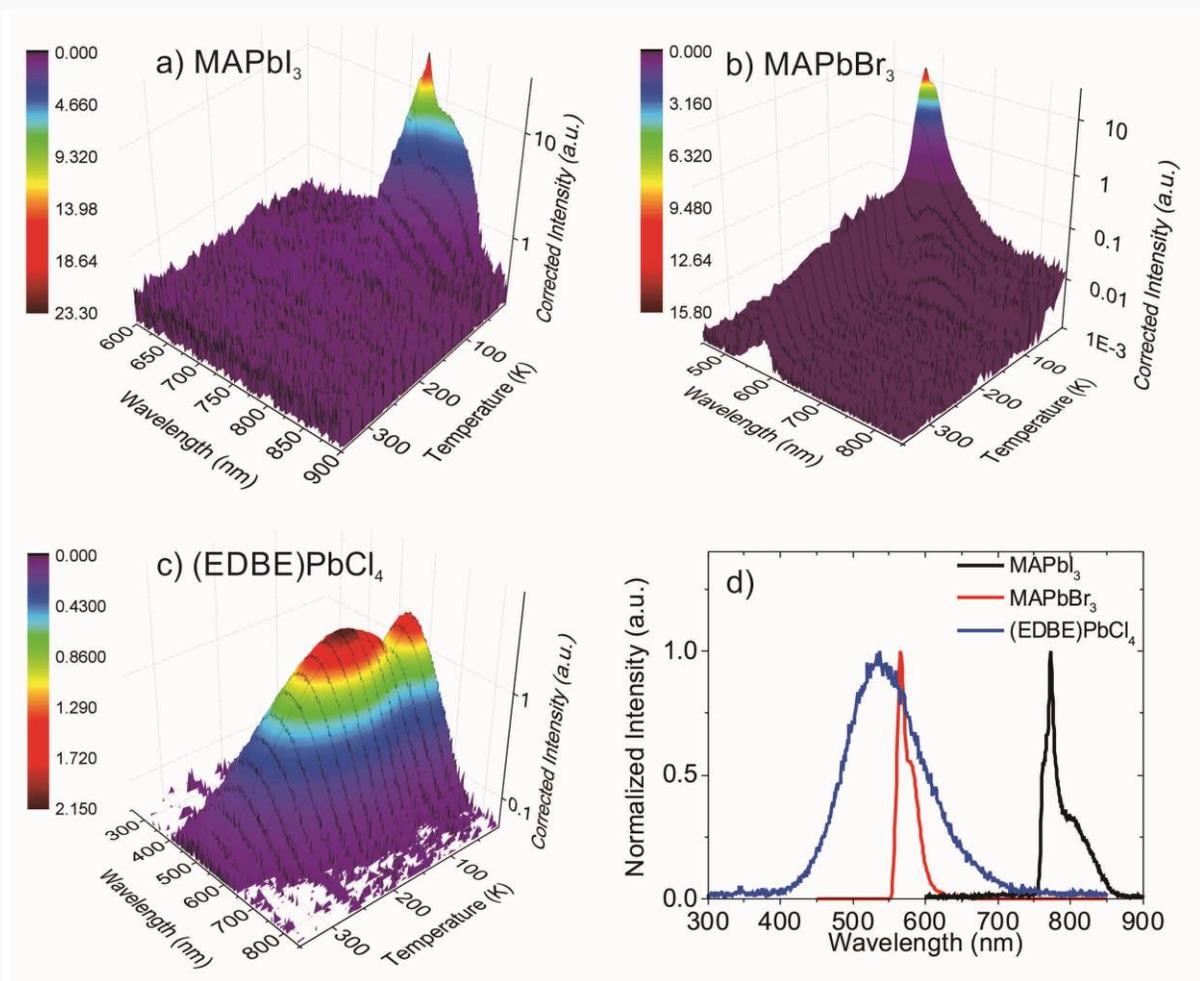

**Figure 5. Temperature dependent luminescence under X-ray excitation.** X-ray excited luminescence spectra (X-ray excited luminescence) of perovskite crystals at various temperatures, from 10 to 350 K: (a) $MAPbI_3$, (b) $MAPbBr_3$, and (c) $(EDBE)PbCl_4$. (d) Comparison of the normalized X-ray excited luminescence spectra at T=10 K.

$MAPbI_3$ (Fig. 5a) and $MAPbBr_3$ (Fig. 5b) show strong dependence of X-ray excited luminescence on temperature, with significantly reduced emission at temperatures above 100 K. At very low temperatures they display distinct emission bands with sharp maxima at 770 nm and 540 nm, respectively (see Fig. 5d for comparison of X-ray excited luminescence spectra recorded at T=10 K). Notably, the X-ray excited luminescence peak at 770 nm, with FWHM of 5 nm, has the same characteristics of coherent light emission previously observed in MAPbI$_3$ [17]. Side bands also appear at 760 and 800 nm in MAPbI$_3$ and at ~590 nm in MAPbBr$_3$, but their origin is still unclear. The X-ray excited luminescence spectrum of (EDBE)PbCl$_4$ (Fig. 5c) consists of a much broader band peaking at ~520 nm, with intensity significantly less sensitive to temperature. As temperature increases, the X-ray excited luminescence intensity first decreases between 10 and 50 K, then increases towards 130 K, and reduces steadily at higher temperatures. In all crystals, the FWHM of X-ray excited luminescence peaks increases with increasing temperature, consistent with the spreading of excited electrons to high vibrational levels [37].

As discussed previously, the light yield of perovskite single crystals estimated from their bandgaps should be >120,000 photons/MeV. From the pulse height spectra of samples excited with 662 keV $\gamma$-ray of Cs$^{137}$ shown in Supplementary Fig. S7, the actual light yield of (EDBE)PbCl$_4$ at room temperature is moderate, with ~9,000 photons/MeV. We note that they are not so many reports about the energy spectra from $\gamma$-ray reported for perovskite scintillator [10] and direct conversion detector [12]. Light yield of (EDBE)PbCl$_4$ is actually similar to that of two-dimensional perovskite crystal PhE-PbBr$_4$ (10,000 photons/MeV) reported previously [31]. The light yields of MAPbBr$_3$ and MAPbI$_3$ at room temperature are much lower, and cannot be extracted from pulse height experiments. Low light yields at room temperature may arise from thermally activated quenching effects. To confirm this hypothesis, we have derived light yields at different temperatures from the integrated intensities of the X-ray excited emission spectra in Fig. 5; considering the small afterglow, we expect the light yield derived from X-ray excited emission spectra to be very similar to that derived from pulse height spectra.

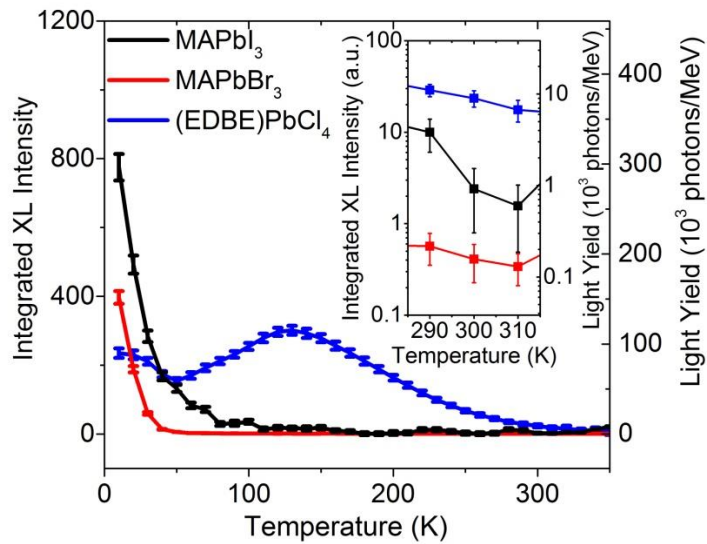

**Figure 6. Temperature dependence of the light yields.** Light yields of MAPbI$_3$, MAPbBr$_3$, and (EDBE)PbCl$_4$ obtained from the integrated X-ray excited luminescence intensities at various temperatures, from 10-350 K. The left axis shows integrated intensity in arbitrary units obtained from the corrected X-ray excited luminescence spectra in Fig. 5, while the right axis exhibits the light yield in absolute units after calibration of the light yield of (EDBE)PbCl$_4$ at 300 K to ~9,000 photons/MeV, as derived independently from its pulse height spectrum. The inset shows details of the curves from 290 to 310 K.

Light yields derived from the integrated X-ray excited luminescence emission intensities of the three halide perovskite crystals as a function of temperature are reported in Fig. 6. We integrated the corrected intensity of X-ray excited luminescence spectra in Fig. 5 and used the light yield of ~9,000 photons/MeV derived from the pulse height spectrum of (EDBE)PbCl$_4$ at 300 K (Supplementary Fig. S7) to calibrate the integrated intensity. For (EDBE)PbCl$_4$, the resulting light yield at 300 K is about ~8% of the maximum at 130 K. Since the light yield is linearly proportional to the photoluminescence quantum efficiency [18] while the efficiency of charge transport to the recombination center is almost unity [24,29], the ratio of 8% is consistent with reported (EDBE)PbCl$_4$ photoluminescence quantum efficiency of less than 10% at room temperature. Light yields of MAPbI$_3$ and MAPbBr$_3$ are <1,000 photons/MeV at room temperature (see inset of Fig. 6) and the light yields at 10 K are 296,000 and 152,000 photons/MeV, respectively. The maximum light yields of MAPbI$_3$ and MAPbBr$_3$ correspond well to the attainable light yields derived from bandgaps of 270,000 and 190,000 photons/MeV, respectively. Unlike (EDBE)PbCl$_4$, the ratio between the light yields at 300 K and those at 10 K of less than 0.5% for MAPbI$_3$ and MAPbBr$_3$ are much smaller than their respective quantum efficiencies of 30% and 10% [17,38]. Additional light-yield-loss in MAPbI$_3$ and MAPbBr$_3$ could be due to non radiative recombination of free electrons and holes within the ionization tracks [2,18].

The larger light yield of (EDBE)PbCl$_4$ at room temperature can be explained by its extremely large exction-binding energy of about 360 meV [29], which is typical of 2D perovskites and gives rise to a pronounced excitonic absorption below the band-edge (see Supplementary Fig. S4). In contrast, 3D perovskites are known for their low exciton binding energy - for MAPbBr$_3$ and MAPbI$_3$ in the range of 2-70 meV [22,23]. Loosely bound excitonic states in 3D perovskites are much more prone to thermal quenching than tightly bound excitons in 2D perovskites like (EDBE)PbCl$_4$, implying that 3D perovskite crystals require much lower operating temperatures than 2D crystals to achieve optimal scintillation performance.

**Conclusions**

Our findings confirm that hybrid lead halide perovskite single crystals are very promising scintillator materials in term of low fabrication costs, low intrinsic trap density, nanosecond fast response, and potentially high light yield. Thermoluminescence measurements have indicated that perovskite crystals have much lower trap density than conventional oxide scintillator materials [19,33]. Low-temperature X-ray excited luminescence measurements have shown that the X-ray luminescence yield can be as high as ~120,000 photons/MeV in (EDBE)PbCl$_4$ at T=130 K, and in excess of 150,000 photons/MeV in MAPbI$_3$ and MAPbBr$_3$ at T=10 K. The wide synthetic versatility of hybrid perovskites allows easy tuning of their scintillation properties: for example, their emission spectra can be controlled by cation or halide substitution to perfectly match the spectral sensitivity of high-quantum-efficiency APD, like in the case of MAPbBr$_3$ and (EDBE)PbCl$_4$. Their emissive properties can be further enhanced through engineering of perovskite structure and dimensionality: while light yield of 3D perovskites MAPbI$_3$ and MAPbBr$_3$ is significantly reduced at room temperature (<1,000 photons/MeV), the 2D perovskite (EDBE)PbCl$_4$ is less affected by thermal quenching (~9,000 photons/MeV at room temperature), thanks to its large exciton binding energy. Given the potential of hybrid lead halide perovskite crystals, further efforts should be made to synthesize new materials for X- and $\gamma$-ray scintillation: for instance, the light yield of perovskite crystals could be further improved by introduction of lanthanide ions, e.g. Ce$^{3+}$ ions, as impurities [18,39], or mixing the halides to modify the bandgap [40], while the optimal operating temperature could be increased through the design of wide band gap 2D perovskite crystals with minimal quenching effects.

**Materials and methods**

*X-ray scintillation measurements*: The main setup used for X-ray excited luminescence and thermoluminescence measurements consists of an Inel XRG3500 X-ray generator (Cu-anode tube, 45 kV / 10 mA), an Acton Research Corporation SpectraPro-500i monochromator, a Hamamatsu R928 PMT, and an APD Cryogenics Inc. closed-cycle helium cooler. The emission spectra were corrected for the transmittance of the monochromator and the quantum efficiency of the PMT. First, we recorded X-ray excited luminescence at various temperatures, between 10 and 350 K. We note that the measurements were carried out from 350 K to 10 K to avoid possible contributions from thermal release of charge carriers to the emission yield. After X-ray excited luminescence measurements, we measured low temperature thermoluminescence and glow curves. Prior to thermoluminescence measurements, the samples were exposed to X-rays for 10 min at 10 K. Thermoluminescence and glow curves were recorded between 10 and 350 K at a heating rate of about 0.14 K/s. Thermoluminescence curves were recorded with the monochromator set to the zeroth order. Photoluminescence spectra were recorded with a commercial spectrofluorometer HORIBA Jobin Yvon Fluorolog-3 spectrofluorometer at room temperature.

*Crystal growth*: Three-dimensional perovskite precursors, MABr and MAI, were synthetized by mixing hydrobromic acid (48% wt in water, Sigma-Aldrich) and hydroiodic acid (57% wt in water, Sigma-Aldrich) with methylamine solution ($CH_3NH_2$, 40% in methanol, Tokyo Chemical Industry, Co., Ltd) in 1:1 molar ratio. The ice-cooled mixture was left under magnetic stirring for 2 h, and the solvent removed with a rotary evaporator. The resulting powders were dissolved in ethanol, crystallized with diethylether for purification repeating the cycle 6 times, and finally dried in vacuum oven at 6 °C for 12 h. For (EDBE)$PbCl_4$(EDBE=2,2'-(ethylenedioxy)bis(ethylammonium)), the organic precursor (EDBE)$Cl_2$ was synthetized in aqueous solution by reaction of 2,2'-(ethylenedioxy)bis(ethylamine) (98%, Sigma Aldrich) with excess of HCl (37% in $H_2O$). The solution was stirred for 4h at room temperature to complete the reaction. A purification process similar to that discussed for MABr and MAI was applied to collect the final white and high purity powders.

For the synthesis of hybrid perovskite crystals, the following inorganic precursors were purchased from Sigma-Aldrich: lead(II) chloride ($PbCl_2$, 99.999%), lead(II) bromide ($PbBr_2$, 99.999%) and lead(II) iodide (PbI$_2$, 99.0%). Crystals of MAPbBr$_3$ were synthetized using inverse temperature crystallization as similarly reported elsewhere [27]. 2 ml of 1M DMF solution of MABr and PbBr$_2$ (1:1 molar ratio) were left overnight on a hotplate (110 °C) without stirring, allowing the precipitation of the perovskite crystals. MAPbI$_3$ were obtained by slow evaporation at room temperature of a saturate *N*, *N*-dimethylformamide (DMF) solution of MAI and PbI$_2$ (1:1 molar ratio). To obtain (EDBE)PbCl$_4$ crystals, a 1M solution of (EDBE)Cl$_2$ and PbCl$_2$ (1:1 molar ratio) in dimethylsulphoxide (DMSO) was prepared by dissolving the precursors at 110 °C on a hotplate. After natural cooling of the solution at room temperature, slow crystallization over a period of 1 month results in the formation of cm-scale white perovskite crystals. The crystallization processes were performed under inert N$_2$ atmosphere. All the obtained crystals were collected from the precursor solutions, washed with diethylether and dried in vacuum overnight.

## Acknowledgements

Research was supported by the Ministry of Education (MOE2013-T2-044 and MOE2011-T3-1-005) and by the National Research Foundation (NRF-CRP14-2014-03) of Singapore. X-ray excited luminescence and thermoluminescence measurements were performed at the National Laboratory for Quantum Technologies (NLTK), Nicolaus Copernicus University, supported by the European Regional Development Fund.

## Author contributions

MDB and CS conceived the idea. DC synthesized the perovskite precursors, prepared, and characterized crystals and films. MDB and WD designed the experiments. KB and WL performed X-ray excited luminescence and thermoluminescence measurements. DC and AB collected absorption and photoluminescence measurements of thin films. WD, MDB, DC and CS analyzed the data. MDB and CS drafted the manuscript. All the authors contributed to interpretation of the results and revision of the manuscript. CS supervised the work. All authors take full responsibility for the content of the paper.

**Additional information**

Supplementary information provided: powder X-ray diffraction measurements of the perovskite crystals to confirm the expected perovskite structures; absorption and photoluminescence measurements of perovskite thin films as a reference for the optical properties of the crystals; calculation of X-ray absorption lengths of the three perovskite crystals, which turn out to be comparable to those of commercial $LaBr_3$ and $LuI_3$ scintillators; time-resolved photoluminescence of single crystals as some decay components are much faster than those in commercial scintillators; pulse height spectra that provide additional details on scintillation properties.

The authors declare that they have no competing financial interests. Reprints and permission information is available online at http://npg.nature.com/reprintsandpermissions/. Correspondence and requests for materials should be addressed to MDB (mbirowosuto@ntu.edu.sg) and CS (csoci@ntu.edu.sg).

**References**
[1] Blasse, G., Scintillator materials, *Chem. Mater.* **6**, 1465-1475 (1994).
[2] Nikl, M., Scintillation detectors for x-rays, *Meas. Sci. Technol.* **17**, R37-R54 (2006).
[3] Street, R. A., Ready, S. E., Van Schuylenbergh, K., Ho, J., Boyce, J. B., Nylen, P., Shah, K., Melekhov, L., and Hermon, H., Comparison of $PbI_2$ and $HgI_2$ for direct detection active matrix x-ray image sensors, *J. Appl. Phys.* **91**, 3345-3355 (2002).

[4] Szeles, C., CdZnTe and CdTe materials for X-ray and gamma ray radiation detector applications, *phys. stat. sol. (b)* **241**, 783-790 (2004).

[5] Kasap, S. and Rowlands, J., Direct-conversion flat-panel X-ray image sensors for digital radiography, *Proc. IEEE* **90**, 591-604 (2002).

[6] Büchele, P., Richter, M., Tedde, S. F., Matt, G. J., Ankah, G. N., Fischer, R., Biele, M., Metzger, W., Lilliu, S., Bikondoa, O., Macdonald, J. E., Brabec, C. J., Kraus, T., Lemmer, U., and Schmidt, O., X-ray imaging with scintillator-sensitized hybrid organic photodetectors, *Nature Photon.* **9**, 843-848 (2015).

[7] Heiss, W. and Brabec, C., X-ray imaging: Perovskites target X-ray detection, *Nature Photon.* **10**, 288-289 (2016).

[8] Tegze, M. and Faigel, G., X-ray holography with atomic resolution, *Nature* **380**, 49-51 (1996).

[9] Rieder, R., Economou, T., Wanke, H., Turkevich, A., Crisp, J., Bruckner, J., Dreibus, G., and McSween, H., The chemical composition of Martian soil and rocks returned by the mobile alpha proton x-ray spectrometer: Preliminary results from the x-ray mode, *Science* **278**, 1771-1774 (1997).

[10] Yakunin, S., Sytnyk, M., Kriegner, D., Shrestha, S., Richter, M., Matt, G. J., Azimi, H., Brabec, C. J., Stangl, J., Kovalenko, M. V., and Heiss, W., Detection of X-ray photons by solution-processed lead halide perovskites, *Nature Photon.* **9**, 444-450 (2015).

[11] Wei, H., Fang, Y., Mulligan, P., Chuirazzi, W., Fang, H.-H., Wang, C., Ecker, B. R., Gao, Y., Loi, M. A., Cao, L., and Huang, J., Sensitive X-ray detectors made of methylammonium lead tribromide perovskite single crystals, *Nature Photon.* **10**, 333-339 (2016).

[12] Yakunin, S., Dirin, D. N., Shynkarenko, Y., Morad, V., Cherniukh, I., Nazarenko, O., Kreil, D., Nauser, T., and Kovalenko, M. V., Detection of gamma photons using solution-grown single crystals of hybrid lead halide perovskites, *Nature Photon.* AOP (2016).

[13] Chin, X.Y., Cortecchia, D., Yin, J., Bruno, A., and Soci, C., Lead iodide perovskite light-emitting field-effect transistor, *Nature Comm.* **6**, 7383-1-7383-9 (2015).

[14] Dou, L., Yang, Y. M., You, J., Hong, Z., Chang, W.H., Li, G., and Yang, Y., Solution-processed hybrid perovskite photodetectors with high detectivity, *Nature Comm.* **5**, 5404-1-5404-6 (2014).

[15] Chondroudis, K. and Mitzi, D. B., Electroluminescence from an organic−inorganic perovskite incorporating a quaterthiophene dye within lead halide perovskite layers, *Chem. Mater.* **11**, 3028-3030 (1999).

[16] Tan, Z.-K., Moghaddam, R. S., Lai, M. L., Docampo, P., Higler, R., Deschler, F., Price, M., Sadhanala, A., Pazos, L. M., Credgington, D., Hanusch, F., Bein, T., Snaith, H. J., and Friend, R. H., Bright light-emitting diodes based on organometal halide perovskite, *Nature Nanotech.* **9**, 687-692 (2014).

[17] Xing, G., Mathews, N., Lim, S. S., Yantara, N., Liu, X., Sabba, D., Grätzel, M., Mhaisalkar, S.and Sum, T. C., Low-temperature solution-processed wavelength-tunable perovskites for lasing, *Nature Mater.* **13**, 476-480 (2014).

[18] Birowosuto, M. D. and Dorenbos, P., Novel γ- and X-ray scintillator research: on the emission wavelength, light yield and time response of $Ce^{3+}$ doped halide scintillators, *phys. stat. sol. (a)* **206**, 9-20 (2009).

[19] Drozdowski, W., Wojtowicz, A. J., Lukasiewicz, T., and Kisielewski, J., Scintillation properties of LuAP and LuYAP crystals activated with Cerium and Molybdenum, *Nucl. Instr. Meth. Phys. Res. A* **562**, 254-261 (2006).

[20] Shibuya, K., Koshimizu, M., Takeoka, Y., and Asai, K., Scintillation properties of $(C_6H_{13}NH_3)(2)PbI_4$: Exciton luminescence of an organic/inorganic multiple quantum well structure compound induced by 2.0 MeV protons, *Nucl. Instr. Meth. Phys. Res. B* **194**, 207-212 (2002).

[21] Kishimoto, S., Shibuya, K., Nishikido, F., Koshimizu, M., Haruki, R., and Yoda, Y., Subnanosecond time-resolved x-ray measurements using an organic-inorganic perovskite scintillator, *Appl. Phys. Lett.* **93**, 261901-1-261901-3 (2008).

[22] Saidaminov, M. I., Abdelhady, A. L., Murali, B., Alarousu, E., Burlakov, V. M., Peng, W., Dursun, I., Wang, L., He, Y., Maculan, G., Goriely, A., Wu, T., Mohammed, O. F., and Bakr, O. M., High-quality bulk hybrid perovskite single crystals within minutes by inverse temperature crystallization, *Nature Comm.* **6**, 7586-1-7586-6 (2015).

[23] Shi, D., Adinolfi, V., Comin, R., Yuan, M., Alarousu, E., Buin, A., Chen, Y., Hoogland, S., Rothenberger, A., Katsiev, K., Losovyj, Y., Zhang, X., Dowben, P. A., Mohammed, O. F., Sargent, E. H., and Bakr, O. M., Low trap-state density and long carrier diffusion in organolead trihalide perovskite single crystals, *Science* **347**, 519-522 (2015).

[24] Dohner, E. R., Jaffe, A., Bradshaw, L. R., and Karunadasa, H. I., Intrinsic white-light emission from layered hybrid perovskites, *J. Am. Chem. Soc.* **136**, 13154-13157 (2014).

[25] van Loef, E. V. D., Dorenbos, P., van Eijk, C. W. E., Krämer, K. and Güdel, H. U., High-energy-resolution scintillator: $Ce^{3+}$ activated $LaBr_3$, *Appl. Phys. Lett.* **79**, 1573-1-1573-3 (2001).

[26] Birowosuto, M. D., Dorenbos, P., van Eijk, C. W. E., Krämer, K. W., and Güdel, H. U., Thermal quenching of $Ce^{3+}$ emission in $PrX_3$ ( X = Cl , Br ) by intervalence charge transfer, *J. Phys. Condens. Matter* **19**, 256209-1-256209-16 (2007).

[27] Birowosuto, M. D., Dorenbos, P.. de Haas, J. T. M., van Eijk, C. W. E., Krämer, K. W., and Güdel, H. U., Optical spectroscopy and luminescence quenching of $LuI_3$: $Ce^{3+}$, *J. Lumin.* **118**, 308-316 (2006).

[28] Birowosuto, M. D., Dorenbos, P., van Eijk, C. W. E., Krämer, K. W., and Güdel, H. U., High-light-output scintillator for photodiode readout: $LuI_3$: $Ce^{3+}$, *J. Appl. Phys.* **99**, 123520-1-123520-4 (2006).

[29] Cortecchia, D., Yin, J., Lova, P., Mhaisalkar, S., Gurzadyan, G. G., Bruno, A., and Soci, C., Polaron self-localization in white-light emitting hybrid perovskites, *arXiv:1603.01284* [cond-mat.mtrl-sci] (2016).

[30] Birowosuto, M. D., Dorenbos, P., van Eijk, C. W. E., Krämer, K. W., and Güdel, H. U., $PrBr_3$: $Ce^{3+}$: A New Fast Lanthanide Trihalide Scintillator, *IEEE Trans. Nucl. Sci.* **53**, 3028-3030 (2006).

[31] van Eijk, C. W. E., de Haas, J. T. M., Rodnyi, P. A., Khodyuk, I. V., Shibuya, K., Nishikido, F., and Koshimizu, M., Scintillation properties of a crystal of $(C_6H_5(CH_2)_2NH_3)_2PbBr_4$, *IEEE Nuclear Science Symposium Conference Record* **2008**, 3525-3528 (2008).

[32] Bartram, R. H., Hamilton, D. S., Kappers, L., and Lempicki, A., Electron traps and transfer efficiency of cerium-doped aluminate scintillators, *J. Lumin.* **75**, 183-192 (1997).

[33] Drozdowski, W., Brylew, K., Witkowski, M. E., Wojtowicz, A. J., Solarz, P., Kamada, K., and Yoshikawa, A., "Studies of light yield as a function of temperature and low temperature thermoluminescence of $Gd_3Al_2Ga_3O_{12}$: Ce scintillator crystals, *Opt. Mater.* **36**, 1665-1669 (2014).

[34] Randall, J., and Wilkins, M., The phosphorescence of various solids, *Proc. Roy. Soc. London A* **184**, 366-407 (1945).

[35] Wojtowicz, A. J., Glodo, J., Drozdowski, W., and Przegietka, K. R., Electron traps and scintillation mechanism in YAlO$_3$: Ce and LuAlO$_3$: Ce scintillators *J. Lumin.* **79**, 275-291 (1998).

[36] Ogorodnikov, I. N. and Poryvai, N. E., Thermoluminescence kinetics of lithium borate crystals, *J. Lumin.* **132**, 1318-1324 (2012).

[37] Liu, C., Qi, Z., Ma, C.-g., Dorenbos, P., Hou, D., Zhang, S., Kuang, X., Zhang, J., and Liang, H., High light yield of Sr$_8$(Si$_4$O$_{12}$)Cl$_8$: Eu$^{2+}$ under x-ray excitation and its temperature-dependent luminescence characteristics *Chem. Mater.* **26**, 3709–3715 (2014).

[38] Sutter-Fella, C. M., Li, Y., Amani, M., Ager III, J. W., Toma, F. M., Yablonovitch, E., Sharp, I. D., and Javey, A., High photoluminescence quantum yield in band gap tunable bromide containing mixed halide perovskites, *Nano Lett.* **16**, 800-806 (2016).

[39] Kolk, E. V. D. and Dorenbos, P., Systematic and material independent variation of electrical, optical, and chemical properties of Ln materials over the Ln series ( Ln ) La, Ce, Pr, .., Lu, *Chem. Mater.* **18**, 3458-3462 (2006).

[40] Birowosuto, M. D., Dorenbos, P., van Eijk, C. W. E., Krämer, K. W., and Güdel, H. U., Ce$^{3+}$ activated LaBr$_{3-x}$I$_x$ : High-light-yield and fast-response mixed halide scintillators, *J. Appl. Phys.* **103**, 103517-1-103517-6 (2008).

# Supplementary Information

## X-ray Diffraction

To characterize the crystal structure of the perovskite crystals, X-ray powder diffraction (XRPD) was performed on perovskite powders obtained from ground crystals on a BRUKER D8 ADVANCE with Bragg-Brentano geometry using Cu K$_\alpha$ radiation ($\lambda$ = 1.54,056 Å), step increment of 0.02° and 1 s of acquisition time. The results are shown in Figs. S1, S2 and S3.

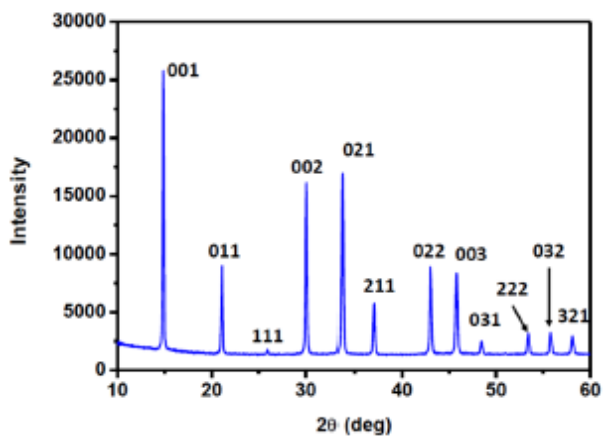

**Figure S1**. XRD pattern of MAPbBr$_3$ powders. The diffractogram is consistent with the perovskite structure having cubic crystal system, space group $Pm3m$ and lattice parameters a = 5.917(1) Å.

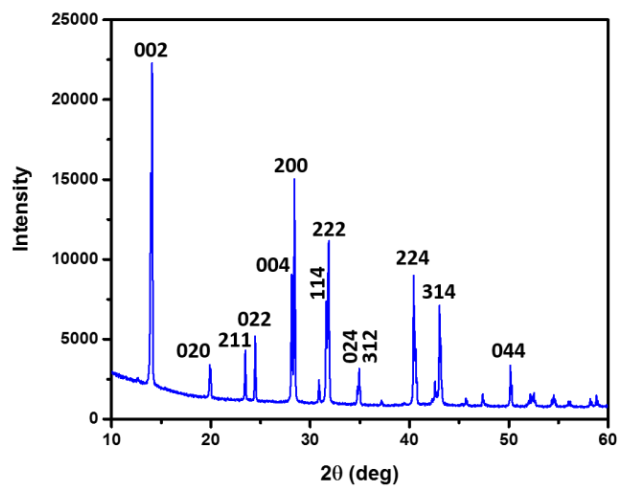

**Figure S2**. XRD pattern of MAPbI$_3$ powders. The diffractogram is consistent with the perovskite structure having tetragonal crystal system, space group $I4/mcm$ and lattice constants a = 8.867(5) Å and b = 12.649(3) Å.

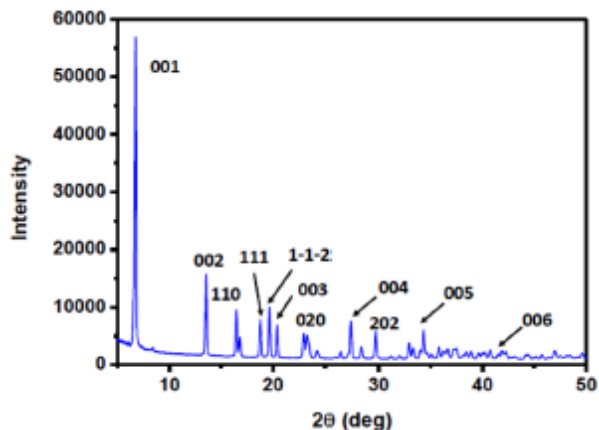

**Figure S3.** XRD pattern of (EDBE)PbCl$_4$ powders. The 00l reflections indicate the formation of the layered structure of the two-dimensional perovskite, in agreement with the previously reported structure (by Dohner *et al*, [1] having monoclinic crystal system, space group C2 and refined lattice parameters a = 7.762(8) Å, b = 7.629(2) Å, c = 13.375(7) Å, $\beta$ = 102.7(2) Å.

**Absorption and Photoluminescence**

Absorption and photoluminescence measurements were performed in order to obtain the energy band gap and confirm the large Stoke shift in two-dimensional perovskite scintillators (Fig. S4). Absorption spectra were recorded by an UV-VIS-NIR spectrophotometer (UV3600, Shimadzu) using a scanning resolution of 0.5 nm. Steady-state photoluminescence spectra were recorded by a Fluorolog-3, (HORIBA Jobin Yvon) spectrofluorometer with wavelength resolution 0.5 nm.

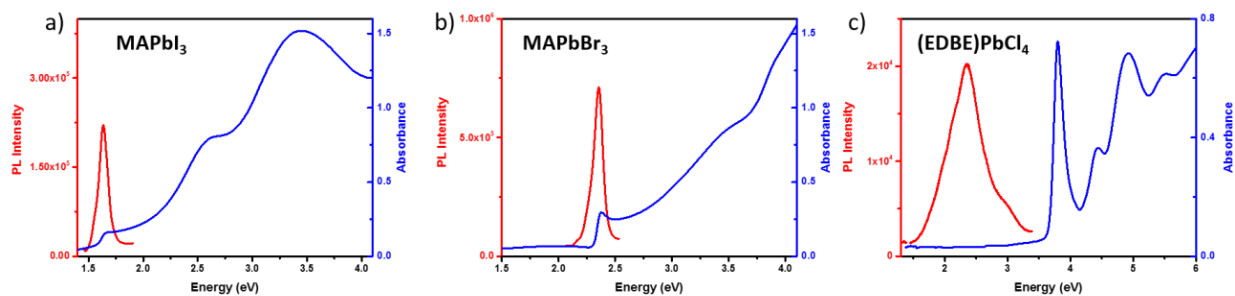

**Figure S4**. Room temperature absorption (blue) and photoluminescence spectrum (red) of a) MAPbI$_3$, b) MAPbBr$_3$, c) (EDBE)PbCl$_4$ (thin films). The photoluminescence peak blue-shifts from 760 nm to 527 nm from MAPbI$_3$ to MAPbBr$_3$, following the corresponding blue-shift of the absorption edges. The absorption spectrum of (EDBE)PbCl$_4$ shows a pronounced excitonic peak at 326 nm and broadband, highly Stoke-shifted photoluminescence peaked at 525 nm.

**Absorption Length**

Three types of the interaction mechanisms for electromagnetic radiation in matter play an important role in the absorption of X- and γ-rays. These are photoelectric absorption, Compton scattering, and pair production. All these processes lead to the partial or complete absorption of the radiation quantum. The absorption length was obtained from formula [2]:

$$l_{abs} \approx \frac{2Z_{eff}}{N_A} \frac{1}{\rho\sigma}$$

where $Z_{eff}$, $N_A$, $\rho$, and $\sigma$ are the effective atomic number, Avogadro number, the mass density, and the absorption cross section for each atomic element. As the cross section was separately characterized for photoelectric, Compton scattering, and pair production, the total absorption length was determined by the inverse sum of the absorption lengths for the three interaction mechanisms mentioned above (Fig. S5).

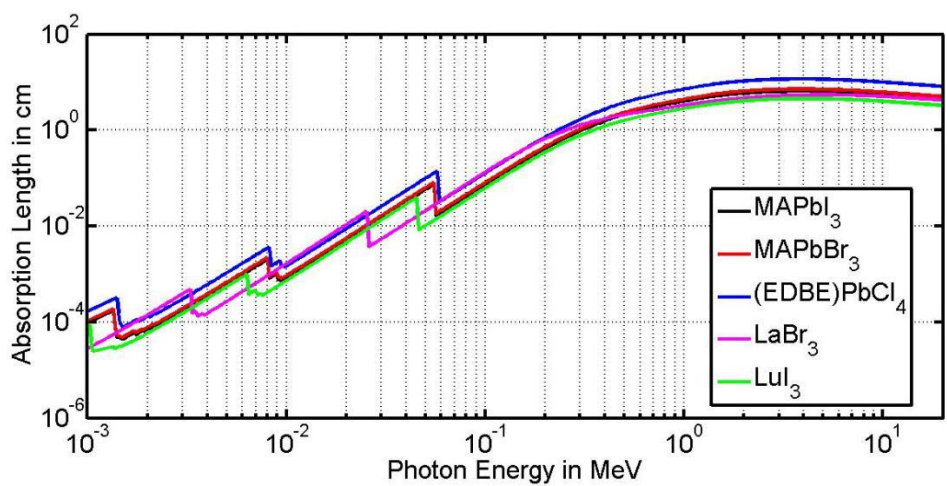

**Figure S5.** Calculated absorption length of perovskite crystals as a function of photon energy, covering the X-ray spectral region. Curves for $Ce^{3+}$ doped $LuI_3$ [2] and $LaBr_3$ [3] scintillators are also added for comparisons.

**Time Resolved Photoluminescence**

The microphotoluminescence setup is based on free space excitation technique with the excitation path and the emission collection from the side, using a VIS-NIR microscope objective (40x, NA=0.65). The MAPbI$_3$, MAPbBr$_3$, and (EDBE)PbCl$_4$ single crystals were excited with 5-MHz-repetition-rate, picosecond-pulse light sources at 640 and 370 nm of Edinburgh laser diodes and at 330 nm of a Picoquant light-emitting diode, respectively. In all cases, the beam spot size was about 2 mm. A silicon-based charge-coupled-device camera was used for imaging. Time-resolved decay curves were obtained using grating Edinburgh Instruments or tunable bandpass filters at 766, 540, and 520 nm for MAPbI$_3$, MAPbBr$_3$ and (EDBE)PbCl$_4$ crystals, respectively. The signal from the Hamamatsu photomultiplier or Micro Photon Devices single-photon avalance photodiode was acquired by a time-correlated single photon counting card.

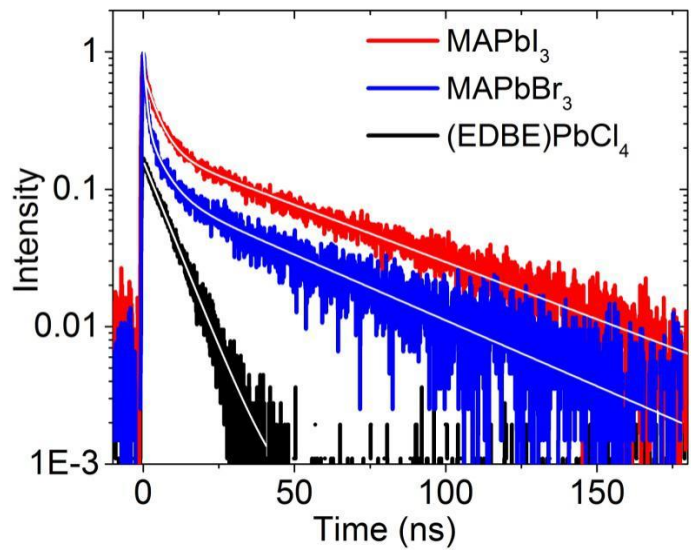

**Figure S6.** Time resolved photoluminescence curves of MAPbI$_3$, MAPbBr$_3$ and (EDBE)PbCl$_4$ single crystals. Excitation and emission wavelengths are reported in the text. All intensities were normalized, and the one of (EDBE)PbCl$_4$ was further divided by a factor of five for clarity. The white lines in the curves are exponential fittings of the data.

The decay curves of MAPbI$_3$, MAPbBr$_3$ and (EDBE)PbCl$_4$ were fitted with double, triple, and single exponential fits, respectively. The resulting decay components of MAPbI$_3$ are 4.3 and 52.2 ns, with contributions of 18 and 82 %, respectively. Those of MAPbBr$_3$ are 0.8, 5.2, and 45.4 ns with contributions of 10, 18 and 72 %, respectively. While the longer decay times are consistent with the values previously reported for these 3D perovskites [4], our instrumental resolution (0.05 ns) allowed to resolve the additional presence of the fast components with decay times < 1 ns. Due to the limited time window, ultralong-lived components (> 300 ns) were not detected. Finally, (EDBE)PbCl$_4$ decay has only one component of 7.9 ns, consistent with the photoluminescence lifetime reported for similar 2D perovskites [1]. Note that all the fast components are below 10 ns, much faster than those of commercial scintillators based on Ce$^{3+}$ doped LuI$_3$ [2] and LaBr$_3$ [3].

**Pulse Height Spectra**

Pulse height spectra were measured at room temperature under 662 keV gamma excitation from a $^{137}$Cs source (no. 30/2010, 210 kBq). The pulsed output signal from a Hamamatsu R2059 photomultiplier was processed by a Canberra 2005 integrating preamplifier, a Canberra 2022 spectroscopy amplifier, and a multichannel analyzer. To improve the light collection efficiency the samples were coupled to the quartz window of the photomultiplier with Viscasil grease and covered with several layers of Teflon tape. Light yield is obtained from the position of the 662 keV photopeak in pulse height spectra both recorded with the photomultiplier and the APD. Using the photomultiplier, the photoeletron yield, expressed in photoelectrons per MeV of absorbed γ-ray energy (phe/MeV), is determined by comparing of the peak position of the 662 keV photopeak to the position of the mean value of the single electron response [2].

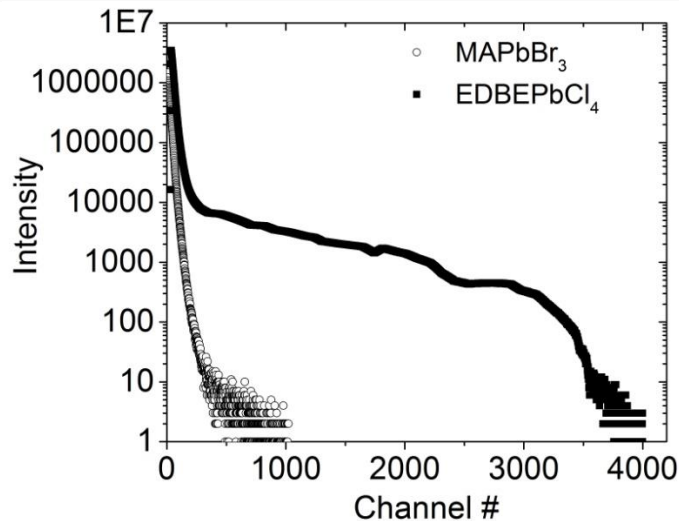

**Figure S7**. Pulse height spectra of perovskite crystals under 662 keV of $^{137}$Cs source with a pulse shaping time of 2 μs. The room temperature light yields derived for (EDBE)PbCl$_4$ and MAPbBr$_3$ are 9,000 and <1,000 photons/MeV, respectively.

## References

[1] Dohner, E. R., Jaffe, A., Bradshaw, L. R., and Karunadasa, H. I., Intrinsic white-light emission from layered hybrid perovskites, *J. Am. Chem. Soc.* **136**, 13154-13157 (2014).

[2] Birowosuto, M. D., Dorenbos, P., van Eijk, C. W. E., Krämer, K. W., and Güdel, H. U., High-light-output scintillator for photodiode readout: LuI$_3$: Ce$^{3+}$, *J. Appl. Phys.* **99**, 123520-1-123520-4 (2006).

[3] van Loef, E. V. D., Dorenbos, P., van Eijk, C. W. E., Krämer, K. and Güdel, H. U., High-energy-resolution scintillator: Ce$^{3+}$ activated LaBr$_3$, *Appl. Phys. Lett.* **79**, 1573-1-1573-3 (2001).

[4] Shi, D., Adinolfi, V., Comin, R., Yuan, M., Alarousu, E., Buin, A., Chen, Y., Hoogland, S., Rothenberger, A., Katsiev, K., Losovyj, Y., Zhang, X., Dowben, P. A., Mohammed, O. F., Sargent, E. H., Bakr, O. M., Low trap-state density and long carrier diffusion in organolead trihalide perovskite single crystals, *Science* **347**, 519 (2015).